\documentclass[a4paper, 10pt, twocolumn]{article}

\usepackage{amsfonts,amssymb,amsmath,amsthm}
\usepackage{graphicx}
\usepackage[footnotesize]{caption}
\usepackage{algorithm}
\usepackage[noend]{algorithmic}
\usepackage{siunitx}
\usepackage{booktabs}
\usepackage{multirow}
\usepackage{subcaption}
\usepackage{hyperref}

\DeclareMathOperator*{\diag}{diag}
\DeclareMathOperator*{\cond}{cond}
\DeclareMathOperator*{\argmin}{argmin}
\newcommand{\norm}[1]{\left\lVert#1\right\rVert}

\usepackage[top=1.1cm, left=1.1cm, right=1.1cm, bottom=1.1cm]{geometry}

\title{Joint deconvolution and blind source separation on the sphere\\ with an application to radio-astronomy}

\author{R. Carloni Gertosio and J. Bobin\\
	\footnotesize IRFU, CEA, Université Paris-Saclay, F-91191 Gif-sur-Yvette, France
}
\date{\empty} % no need for a date

\renewenvironment{abstract}{\bf\small {\em\ Abstract---}}{}

\begin{document}
	
	\setlength{\abovedisplayskip}{-1pt}
	\setlength{\belowdisplayskip}{-1pt}
	\setlength{\belowcaptionskip}{-10pt}
	
	\maketitle
	
	\begin{abstract} 
	Blind source separation is one of the major analysis tool to extract relevant information from multichannel data. While being central, joint deconvolution and blind source separation (DBSS) methods are scarce. To that purpose, a DBSS algorithm coined SDecGMCA is proposed. It is designed to process data sampled on the sphere, allowing large-field data analysis in radio-astronomy. 
	\end{abstract}

	\vspace{-0.9em}
	\subsection*{Context}
	\vspace{-0.3em}
	With the forthcoming large-scale radio-telescope, such the Square Kilometer Array (SKA)\footnote{https://www.skatelescope.org/}, standard blind source separation algorithms are faced with a key bottleneck: accounting for instrumental response calls for jointly tackling a separation and deconvolution problem. Additionally, dedicated methods must be designed to specifically address the spherical data that large arrays of radio-telescopes will produce. In this context, multichannel data are considered where $\{\mathbf{Y}_\nu \in \mathbb{R}^{N_p}, \nu \in [1,N_c]\}$ are a set of $N_c$ multiwavelength spherical channels. Each channel $\nu$ is deteriorated by an isotropic convolution kernel $\mathbf{H}_\nu \in \mathbb{R}^{N_p}$. Each channel is corrupted with additive Gaussian noise, and modeled as the linear combination of $N_s$ sources, thus leading at channel $\nu$: $\mathbf{Y}_\nu = \mathbf{H}_\nu * (\mathbf{A}_\nu \mathbf{S}) + \mathbf{N}_\nu$, with $*$ the convolution product on the sphere, $\mathbf{S} \in \mathbb{R}^{N_s\times N_p}$ the sources, $\mathbf{A_\nu}$ the $\nu$\textsuperscript{th} row of the mixing matrix $\mathbf{A}\in \mathbb{R}^{N_c\times N_s}$ and $\mathbf{N_\nu} \in \mathbb{R}^{N_p}$ the noise. The equation can be simplified in the spherical harmonics domain for each harmonic coefficient $(l,m)$ and for all channels: 
	$\mathbf{\hat{Y}}^{l,m} = \diag({\mathbf{\hat{H}}^{l}}) \mathbf{A} \mathbf{\hat{S}}^{l,m} + \mathbf{\hat{N}}^{l,m}$. The convolution kernel being isotropic, it only depends on the parameter $l$. In this work, the Healpix pixelisation of the sphere is used \cite{pixel:healpix}.
	
	\underline{\bf Contributions:} In the present paper, we extend the algorithm DecGMCA \cite{Jiang_2017} to tackle such problems from spherical data. Based on a projected alternate least-square minimization, the joint deconvolution/separation procedure calls for extra regularizations to deal a naturally ill-conditioned if not ill-posed problem. Beyond a mere extension, we further introduce several regularization strategies, which significantly improve the separation quality.
	
	\vspace{-0.9em}
	\subsection*{Methodology}
	\vspace{-0.3em}
	In the scope of the joint DBSS problem, the objective is to estimate $\mathbf{A}$ and $\mathbf{S}$ from $\mathbf{Y}$, knowing $\mathbf{H}$ and the level of the noise $\mathbf{N}$. The problem amounts to minimizing an objective function:
	\begin{equation}
	\label{eq:jointdecsep}
	\argmin_{\mathbf{A}\in\mathcal{O},\mathbf{S}} \text{~~} \dfrac{1}{2} \sum_{l,m} \norm{\mathbf{\hat{Y}}^{l,m} - \diag\left({\mathbf{\hat{H}}^{l}}\right) \mathbf{A} \mathbf{\hat{S}}^{l,m}}^2_2 
	+   \norm{\mathbf{\Lambda} \odot \left( \mathbf{S} \mathbf{\Phi}^T\right)}_{\ell_1}
	\end{equation}
	The sources are assumed to be sparse in a representation $\mathbf{\Phi}$, hence the $\ell_1$-penalization of sparsity parameters $\mathbf{\Lambda}$. To mitigate the scale indeterminacy of the product $\mathbf{A \hat{S}}$, the columns of $\mathbf{A}$ are enforced to be on the $\ell_2$-hypersphere or oblique ensemble $\mathcal{O}$.\\
	The problem in Eq.\eqref{eq:jointdecsep} is not convex but multiconvex, which calls for an alternate minimization according to each variable $\bf A$ and $\bf S$. However, traditional proximal algorithms such as the BCD \cite{Xu_13_BlockCoordinateDescent} or the PALM \cite{Bolte_13_Proximalalternatinglinearized} generally exhibit a clear lack of robustness with respect to the often spurious local critical points of the above cost function. Projected alternate least-squares (pALS) \cite{Paatero_94_Positivematrixfactorization} has long been advocated as allowing for more robust minimization schemes \cite{Gillis_12_AcceleratedMultiplicativeUpdates,KervazoPALM19}. Furthermore, pALS allows for simple and robust heuristics to fix the sparse regularization parameters $\Lambda$ \cite{KervazoPALM19}. Hence, and following the architecture of DecGMCA \cite{Jiang_2017}, the proposed algorithm will build upon a sparsity-enforcing pALS, which iterates are the following:\\
	\textbf{$\bullet$ Estimation of $\mathbf{A}$ with $\mathbf{S}$ fixed:} Solving the least-square problem yields ${}\mathbf{A}_\nu = ( \sum_{l,m} \mathbf{\hat{Y}}_\nu^{l,m} \mathbf{\hat{H}}_\nu^{l} \mathbf{\hat{S}}^{l,m\,\dagger} ) ( \sum_{l,m} \mathbf{\hat{H}}_\nu^{l\,2} \mathbf{\hat{S}}^{l,m} \mathbf{\hat{S}}^{l,m\,\dagger})^{-1}$. Since the number of frequencies is much greater than the number of sources $N_s$, the matrix $( \sum_{l,m} \mathbf{\hat{H}}_\nu^{l\,2} \mathbf{\hat{S}}^{l,m} \mathbf{\hat{S}}^{l,m\,\dagger})$ is well conditioned and safe to invert. The solution is then projected on the multidimensional $\ell_2$-hypersphere $\mathcal{O}$.\\
	\textbf{$\bullet$ Estimation of $\mathbf{S}$ with $\mathbf{A}$ fixed:} The quadratic term of the joint deconvolution/separation problem is likely ill-conditioned, if not ill-posed. To alleviate this problem, an extra Tikhonov regularization is required, which turns the following additional regularization term to Eq.\eqref{eq:jointdecsep}:  $\dfrac{1}{2} \sum_{l,m,n} \varepsilon_{n,l}  \left| \mathbf{\hat{S}}^{l,m}_n \right|^2$. $\{\varepsilon_{n,l}\}$ are the regularization coefficients, which depend on the frequency $l$ and on the source $n$. In \cite{Jiang_2017}, these parameters were fixed to an {\it ad hoc} small value ({\it e.g. $10^{-3}$}). However, these parameters largely impact the quality of the separation. In the sequel, we investigate different strategies allowing more efficient and adaptive way of tuning these key parameters.\\
	Solving the newly formed quadratic term yields $\mathbf{\hat{S}}^{l,m} = ( \mathbf{M}[l] + \diag_{n} \left(\varepsilon_{n,l}\right))^{-1} \mathbf{A}^T \diag({\mathbf{\hat{H}}^{l}}) \mathbf{\hat{Y}}^{l,m}$, with $\mathbf{M}[l] = \mathbf{A}^T \diag(\mathbf{\hat{H}}^{l})^2 \mathbf{A}$. Four strategies that reduce the choice of the parameters to a single one, called the regularization hyperparameter and denoted $c$, are considered:
	\vspace{-0.5em}
	\begin{itemize}
		\setlength\itemsep{-0.25em}
		\item \textbf{Strategy \#1} (naive strategy): the regularization parameters are chosen independently of the frequency $l$ and the source $n$: $\varepsilon_{n,l} = c $. 
		\item \textbf{Strategy \#2} (strategy used in DecGMCA \cite{Jiang_2017}): $\varepsilon_{n,l} = c\, \lambda_{\text{max}}(\mathbf{M}[l])$, where $\lambda_{\text{max}}(\cdot)$ returns the greatest eigenvalue. 
		\item \textbf{Strategy \#3}: $ \varepsilon_{n,l} = \max\left(0, c-\frac{\lambda_{\text{min}}(\mathbf{M}[l])}{\lambda_{\text{min}}(\mathbf{A}^T\mathbf{A})}\right)$, where $\lambda_{\text{min}}(\cdot)$ returns the smallest eigenvalue. This strategy allows to limit the noise amplification to $c\,\lambda_{\text{min}}(\mathbf{A}^T\mathbf{A})$.
		\item \textbf{Strategy \#4} (SNR strategy): $\varepsilon_{n,l} = c/\text{SNR}_n[l] = c\, \sigma^2_{\mathbf{\hat{N}}}/\sigma^2_{\mathbf{\hat{S}}_n}\![l]$, where $\{\sigma^2_{\mathbf{\hat{S}}_n}\![l]\}$ and $\sigma^2_{\mathbf{\hat{N}}}$ are the angular power spectra of the sources and the noise, respectively. This strategy, which supposes to know the angular power spectra of the sources, is reminiscent of a Wiener deconvolution filter.
	\end{itemize}
	\vspace{-0.5em}
	The solution is then soft-thresholded in the transformed domain $\mathbf{\Phi}$. A $\ell_1$-reweighting strategy is implemented \cite{CandesIRL108} to adapt the thresholds to the pixel values, thus reducing the bias introduced by the soft-thresholding and improving the separation performances. The choice of the corresponding sparse regularization parameter $\Lambda$ follows \cite{KervazoPALM19}.\\
	SDecGMCA is initialized using Principal Component Analysis. During the first iterations, regularization strategy \#3 is used; when the estimated sources converge, the strategy is switched to \#4. The first stage allows to have a first estimation of the sources, whose angular power spectra are close enough to the ground-truth ones (\textbf{warm-up}). The second stage allows to refine the results, by using a more precise regularization strategy (\textbf{refinement}); more specifically, the regularization parameters are calculated with the angular power spectra of the sources estimated at last iteration. SDecGMCA needs to be provided the regularization hyperparameters at warm-up $c_{wu}$ and refinement $c_{ref}$. As proposed in DecGMCA, a decrease of the warm-up regularization hyperparameter is implemented to improve the robustness of the algorithm.
	\vspace{-0.5em}
	\begin{algorithm} 
		\caption{SDecGMCA}
		\begin{algorithmic}
			\STATE Regularization strategy selection: \#3 (\textbf{warm-up})
			\FOR{$i=1, ..., i_{max}$}
			\STATE \textit{(1) Estimate $\mathbf{S}$ with $\mathbf{A}$ fixed}
			\STATE  Tikhonov-penalized least-square update of $\mathbf{\hat{S}}$
			\STATE  Soft-thresholding of $\mathbf{S} \mathbf{\Phi}^T$
			\STATE \textit{(2) Estimate $\mathbf{A}$ with $\mathbf{S}$ fixed}
			\STATE Least-square update of $\mathbf{A}$
			\STATE Projection of $\mathbf{A}$ on $\mathcal{O}$	
			\IF{$\mathbf{S}$ has converged}
			\STATE Regularization strategy selection: \#4 (\textbf{refinement})
			\ENDIF
			\ENDFOR
		\end{algorithmic}
		\label{algo}
	\end{algorithm}
	\vspace{-2em}
	\subsection*{Numerical experiments}
	\vspace{-0.5em}
	We set the Healpix parameters $n_{side}=128$ and $l_{max}=384$. We generate toy example sources, which are sparse in the spherical starlet domain \cite{Starck05} and band-limited to $l_{max}/6 = 64$. We take $N_s = 4$, $N_c = 8$ and $\cond(\mathbf{A}) = 2$. The convolution kernels are Gaussian, with resolutions evenly spread between the minimum resolution $l_{max}/8=48$ and $l_{max}$. The overall $\text{SNR}$ is 10 dB. The channels are unmixed and deconvolved at the resolution of the best-resolved channel. This amounts to replacing $\mathbf{\hat{H}}_\nu^l$ by $\mathbf{\hat{H}}_\nu^l/\mathbf{\hat{H}}_{\nu_b}^l$, where $\nu_b$ is the best-resolved channel number. The performance metrics employed to assess the results are: (i) the normalized mean square error $\text{NMSE} = {||\mathbf{{H}}_{\nu_b}*\mathbf{S}^*-\mathbf{S}||_\text{F}}/||\mathbf{{H}}_{\nu_b}*\mathbf{S}^*||_\text{F}$, with $\mathbf{S}^*$ the ground truth sources and $\mathbf{S}$ the estimated sources, (ii) the mixing matrix criterion \cite{AMCA15} $\mathcal{C}_{\mathbf{A}} = \text{mean}(\mathbf{A}^T\mathbf{A}^*-\mathbf{I})$, with $\mathbf{A}^*$ the ground-truth mixing matrix and $\mathbf{A}$ the estimated mixing matrix.
	
	In order to compare the impact of the 4 regularization strategies, non-blind estimations of $\mathbf{S}$ are performed on a wide range of SNR, $N_c$, $\cond(\mathbf{A})$ and channel resolutions; the results are reported in Table \ref{strat_comp}. Unsurprisingly, strategy \#4 clearly provides the best reconstruction qualities. Among the other strategies, that do not assume the sources to be known, strategy \#3 achieves better results. It is mostly thanks to the non-linear max operator, which allows to keep the lower frequencies unbiased, where most of the sources energy is located. Strategy \#2 gives poor results; indeed, it biases more significantly the lower frequencies than the higher ones.
	\begin{table}
		\centering
		\begin{footnotesize}
			\begin{tabular}{@{}lllll@{}}
				\toprule
				\multirow{2}{*}{Varying parameter} & \multirow{2}{*}{Range} & \multicolumn{3}{c}{Regularization strategy} \\
				\cmidrule{3-5}
				& & \#1        & \#2       & \#3       \\
				\midrule
				SNR (dB)                         & $-20$ to 40 & -9.14 &  -9.73 & -3.63    \\
				$N_c$ & 4 to 25         & -9.84 &  -10.26 & -4.57  \\
				$\cond(\mathbf{A})$       & 1.5 to 14      & -7.33 &  -7.85 & -1.75  \\
				Minimum resolution ($l$) &  2 to 350    & -8.94  &  -9.17 & -5.20 \\ 
				\bottomrule
			\end{tabular}
		\end{footnotesize}
		\caption{Mean NMSE degradation in dB, over 100 realizations, compared to strategy \#4, of the optimized non-blind deconvolution and separation problem. For each point, the equivalent of the finale update of $\mathbf{S}$ in SDecGMCA is performed with the optimal hyperparameter and $\mathbf{A^*}$. For strategy \#4, the regularization parameters are calculated with the angular power spectra of $\mathbf{S^*}$.}
		\label{strat_comp}
	\end{table}
	
\begin{figure}[H]
		\begin{subfigure}{.495\linewidth}
			\centering
			\includegraphics[width=.9\linewidth]{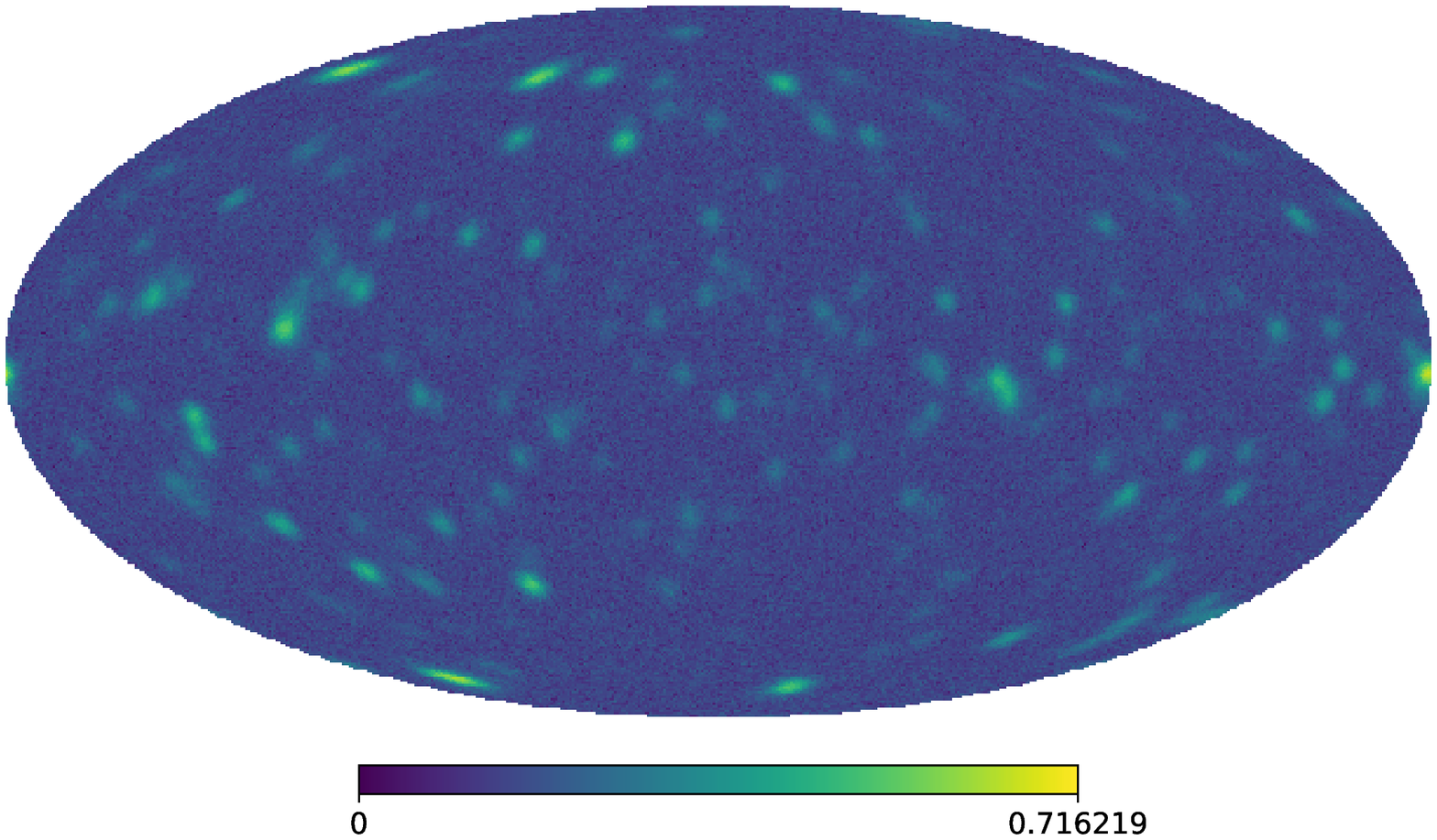}
			\caption{Worse-resolved channel $\mathbf{Y}_1$}
		\end{subfigure}
		\begin{subfigure}{.495\linewidth}
			\centering
			\includegraphics[width=.9\linewidth]{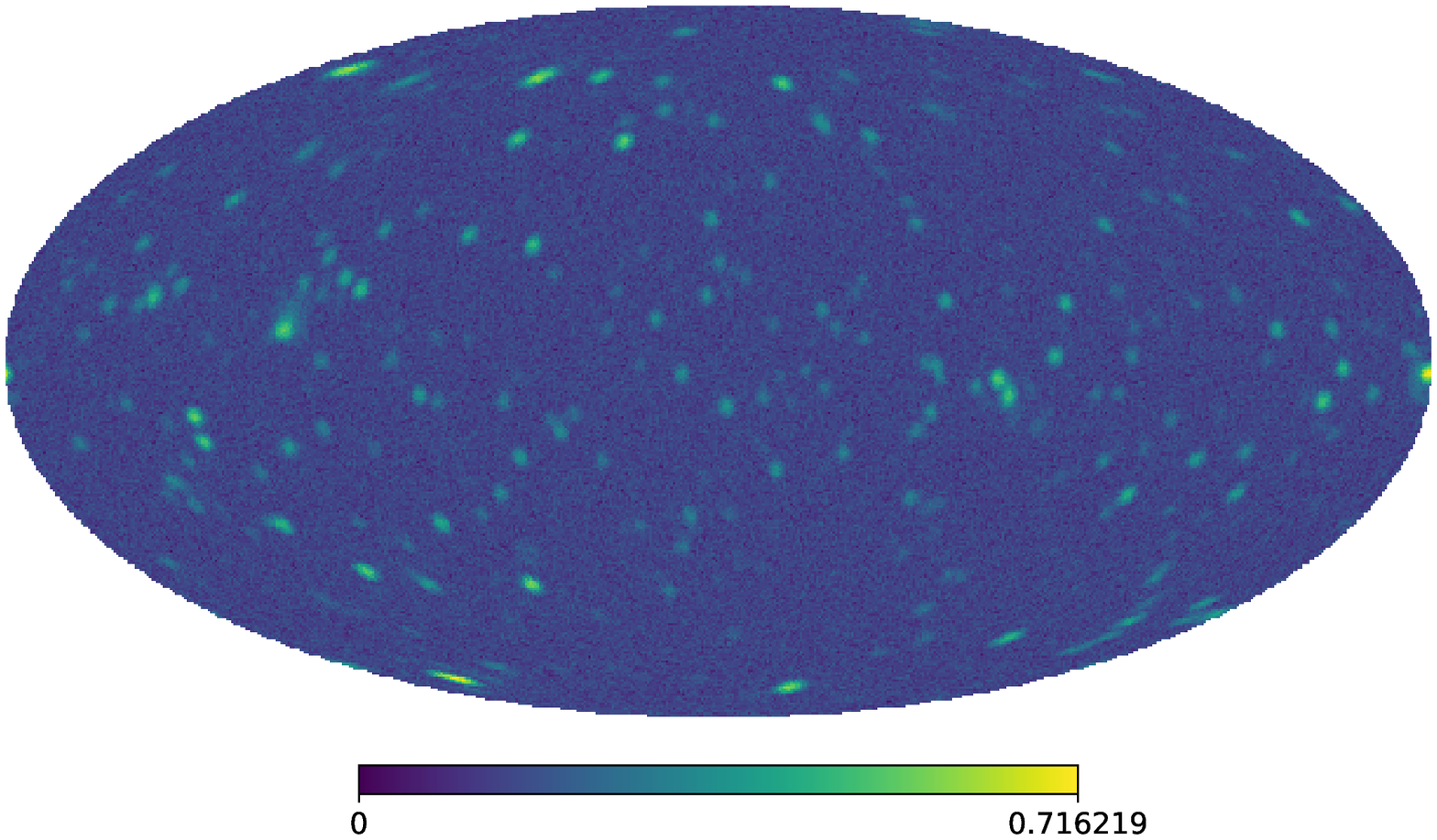}
			\caption{Best-resolved channel $\mathbf{Y}_8$}
		\end{subfigure}
		\begin{subfigure}{.495\linewidth}
			\centering
			\includegraphics[width=.9\linewidth]{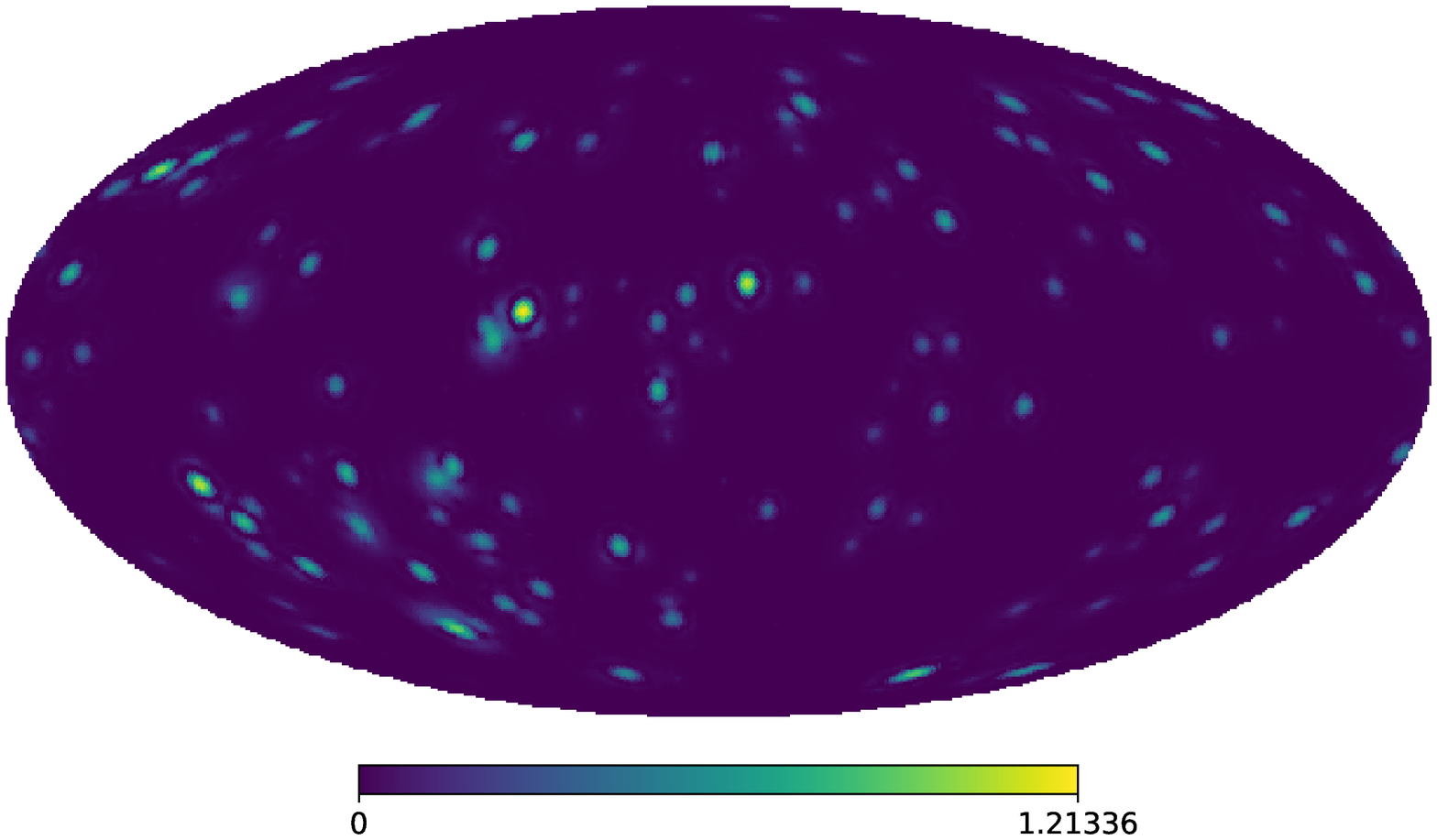}
			\caption{Estimated source $\mathbf{S}_1$}
		\end{subfigure}
		\begin{subfigure}{.495\linewidth}
			\centering
			\includegraphics[width=.9\linewidth]{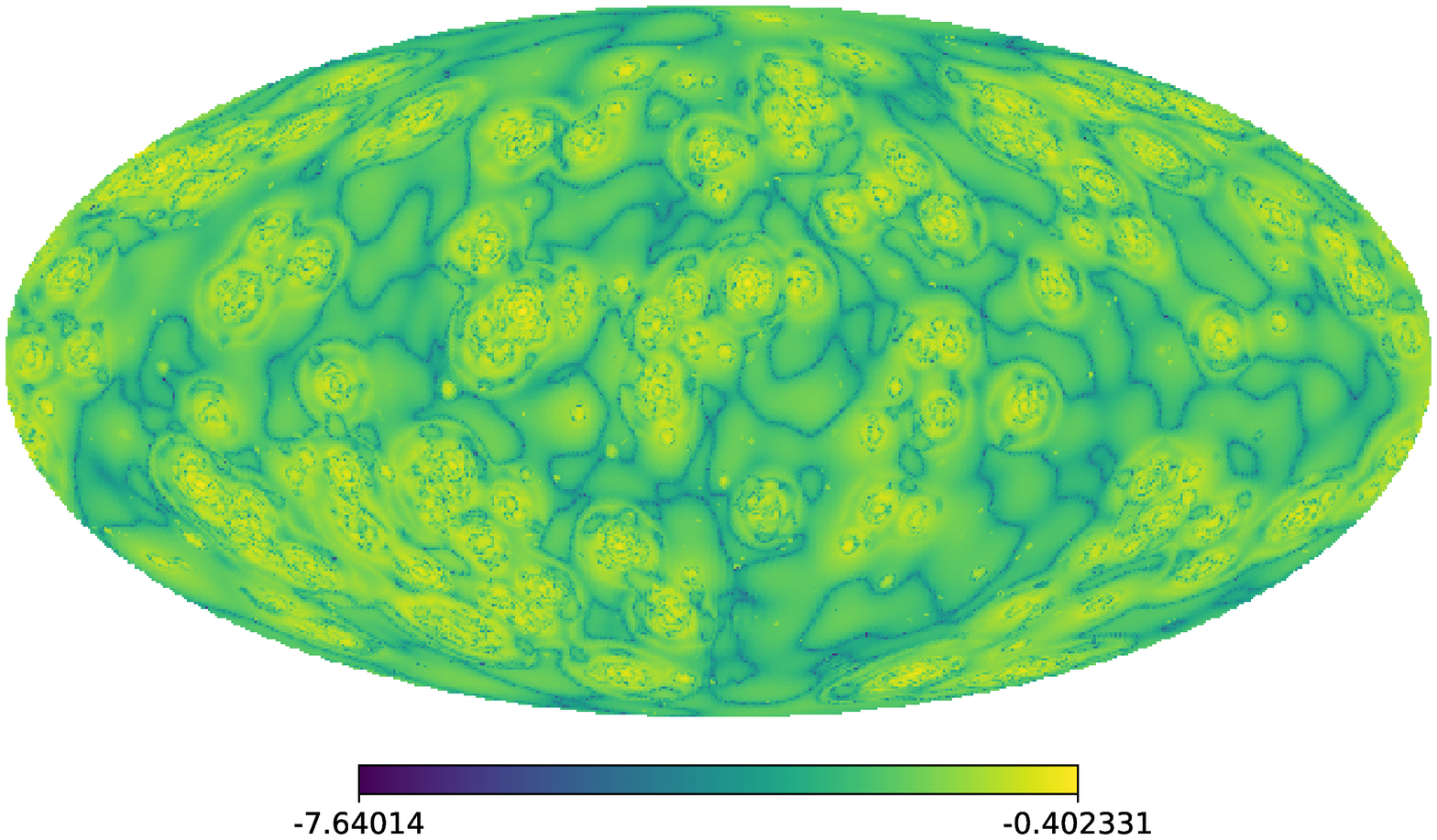}
			\caption{Absolute error $\log_{10}(|\mathbf{H}_8 * \mathbf{S}_1^*-\mathbf{S}_1|)$}
			\label{E0}
		\end{subfigure}
		\vspace{0.2em}
		\caption{DBSS example with SDecGMCA (arbitrary unit, logarithmic scale)}
	\end{figure}

	Table \ref{meanNMSE} shows the mean NMSE of SDecGMCA as a function of the regularization hyperparameters. The choice of $c_{wu}$ has little impact on the NMSE. On the contrary, the selection of $c_{ref}$ is more critical. However, in a range of one order of magnitude around the optimal hyperparameter, the NMSE loss is contained. It is noted that the reconstruction errors are dominated by the deconvolution artifacts (see example Figure \ref{E0}). The mean $\mathcal{C}_{\mathbf{A}}$ varies between 22.07 and 25.60 dB for the same ranges considered in Table \ref{meanNMSE}. Therefore, both $c_{wu}$ and $c_{ref}$ have little impact on the quality of the estimation of $\mathbf{A}$. 
	\begin{table}
		\centering
		\begin{footnotesize}
			\begin{tabular}{@{}lllll@{}} 
				\toprule
				& & \multicolumn{3}{c}{$c_{wu}$ ($\times\, {c_{wu}}_{opt}$)} \\ 
				\cmidrule(r){3-5}
				& & $10^{0.5} \rightarrow 10^{-0.5}$ & $\mathbf{10^{1} \rightarrow 10^{0}}$ & $10^{1.5} \rightarrow 10^{0.5}$ \\ 
				\midrule
				\multirow{5}{*}{\shortstack[c]{$c_{ref}$\\($\times\, {c_{ref}}_{opt}$)}} & $10^{-1} $ & 22.99 & 23.00 & 23.06 \\
				& $10^{-0.5} $ & 24.58 & 24.58 & 24.59 \\
				& $\mathbf{10^{-0}} $    & $\mathbf{24.79}$ & 24.79 & 24.65 \\
				& $10^{0.5} $  & 22.83 & 22.82 & 22.44 \\
				& $10^{1} $     & 18.35 & 18.34 & \textit{18.10} \\
				\bottomrule
			\end{tabular}
		\end{footnotesize}
		\caption{Mean NMSE in dB, over 100 realizations, as a function of $c_{wu}$ and $c_{ref}$. These are given as multiples of ${c_{wu}}_{opt}$ and ${c_{ref}}_{opt}$, which are the mean optimal hyperparameters for the non-blind problem. It is noted that the mean best NMSE of the non-blind problem is 25.74 dB (upper-bound for the blind problem).} 
		\label{meanNMSE}
	\end{table}
	
	SDecGMCA is finally compared to an optimized version of DecGMCA (SDecGMCA with strategy \#2). The results are reported in Table \ref{comparison}. SDecGMCA performs a significant gain in NMSE and a moderate increase in $\mathcal{C}_{\mathbf{A}}$. %The overregularization which occurs in optimized DecGMCA favors the lower frequencies, hence the high $\mathcal{C}_{\mathbf{A}}$, but penalizes the higher frequencies, hence the poor NMSE. 
	SDecGMCA is also compared to two non-deconvolving BSS algorithms. For the latter, the data are deteriorated to a common resolution (the worse one) beforehand. They achieve poor results; indeed, crucial information is lost when the data are deteriorated.
	\begin{table}[H]
		\centering
		\begin{footnotesize}
			\begin{tabular}{@{}llll@{}} 
				\toprule
				& \multirow{2}{*}{$\mathcal{C}_{\mathbf{A}}$ (dB)} & \multicolumn{2}{c}{NMSE (dB)} \\
				\cmidrule(r){3-4}
				& & Worse resolution & Best resolution \\ 
				\midrule
				SDecGMCA                & \textbf{24.81} & \textbf{27.08} & \textbf{24.79}  \\
				Optimized DecGMCA & 23.01 & 20.94 & 15.03 \\
				GMCA                       & 21.98  & 19.35 & N/A \\
				HALS \cite{hals_code}  & 8.17 & 5.83  & N/A \\
				\bottomrule
			\end{tabular}
		\end{footnotesize}
		\caption{Mean performance metrics, over 100 realizations, achieved by different algorithms.  To calculate the worse resolution NMSE for SDecGMCA and optimized DecGMCA, the estimated sources are deteriorated once the algorithm is completed.}
		\label{comparison}
	\end{table}
	
	\vspace{-1em}
	\subsection*{Conclusion}
	\vspace{-0.5em}
	We proposed an enhanced version of DecGMCA, coined SDecGMCA, extended for spherical data. We investigated in particular the regularization and proposed better suited regularization strategies. The results showed that SDecGMCA clearly outperformed DecGMCA. During the workshop, results on realistic simulation data will be presented. 
	
	\newpage
	
	%%% You can make the bibliography smaller
	\bibliographystyle{plain}
	\bibliography{biblio}

\begin{thebibliography}{10}

\bibitem{AMCA15}
J.~{Bobin}, J.~Rapin, J.-L. {Starck}, and A.~Larue.
\newblock Sparsity and adaptivity for the blind separation of partially
  correlated sources.
\newblock {\em IEEE Transactions on Signal Processing}, 63(5):1199--1213, 2015.

\bibitem{Bolte_13_Proximalalternatinglinearized}
J.~Bolte, S.~Sabach, and M.~Teboulle.
\newblock {P}roximal alternating linearized minimization for nonconvex and
  nonsmooth problems.
\newblock {\em Mathematical Programming}, pages 1--36, 2013.

\bibitem{CandesIRL108}
E.~J. Candes, M.~B. Wakin, and S.~P. Boyd.
\newblock Enhancing sparsity by reweighted {L1} minimization.
\newblock {\em Journal of Fourier Analysis and Applications}, 14(5), 2008.
\newblock 877--905.

\bibitem{hals_code}
M.~Chan.
\newblock Hyperspectral image unmixing program code.
\newblock Available on \url{https://github.com/MichelsonChan/hyperspec_unmix},
  last access on 2020-02-13.

\bibitem{Gillis_12_AcceleratedMultiplicativeUpdates}
N.~Gillis and F.~Glineur.
\newblock {A}ccelerated {M}ultiplicative {U}pdates and {H}ierarchical {ALS}
  {A}lgorithms for {N}onnegative {M}atrix {F}actorization.
\newblock {\em Neural Computation}, 24(4):1085--1105, 2012.

\bibitem{pixel:healpix}
K.~M. {G{o}rski}, E.~{Hivon}, A.~J. {Banday}, B.~D. {Wandelt}, F.~K. {Hansen},
  M.~{Reinecke}, and M.~{Bartelmann}.
\newblock {HEALP}ix: A framework for high-resolution discretization and fast
  analysis of data distributed on the sphere.
\newblock {\em Astrophysical Journal}, 622, April 2005.
\newblock 759--771.

\bibitem{Jiang_2017}
M.~Jiang, J.~{Bobin}, and {Starck, J.-L.}
\newblock Joint multichannel deconvolution and blind source separation.
\newblock {\em SIAM Journal on Imaging Sciences}, 10(4):1997--2021, Jan 2017.

\bibitem{KervazoPALM19}
C.~Kervazo, J.~Bobin, C.~Chenot, and F.~Sureau.
\newblock Use of palm for $\ell_1$ sparse matrix factorization: Difficulty and
  rationalization of an heuristic approach.
\newblock {\em Digital Signal Processing}, 97, February 2020.

\bibitem{Paatero_94_Positivematrixfactorization}
P.~Paatero and U.~Tapper.
\newblock {P}ositive matrix factorization: {A} non-negative factor model with
  optimal utilization of error estimates of data values.
\newblock {\em Environmetrics}, 5(2):111--126, 1994.

\bibitem{Starck05}
J.-L. Starck, Y.~Moudden, Abrial P., and M.~Nguyen.
\newblock Wavelets, ridgelets and curvelets on the sphere.
\newblock {\em Astronomy and Astrophysics}, 446(3):1191--1204, 2006.

\bibitem{Xu_13_BlockCoordinateDescent}
Y.~Xu and W.~Yin.
\newblock {A} {B}lock {C}oordinate {D}escent {M}ethod for {R}egularized
  {M}ulticonvex {O}ptimization with {A}pplications to {N}onnegative {T}ensor
  {F}actorization and {C}ompletion.
\newblock {\em SIAM Journal on Imaging Sciences}, 6(3):1758--1789, 2013.

\end{thebibliography}

\end{document}